# Manipulating electronic states at oxide interfaces using focused micro X-rays from standard lab-sources

Nicola Poccia[1,2,a)], Alessandro Ricci[2,3], Francesco Coneri[1], Martin Stehno[1], Gaetano Campi[4], Nicola Demitri[5], Giorgio Bais[5], X. Renshaw Wang[1,b)], H. Hilgenkamp[1]

[1]MESA+ Institute for Nanotechnology, University of Twente, P. O. Box 217, 7500AE Enschede, The Netherlands.
[2]RICMASS Rome International Center for Materials Science Superstripes, via dei Sabelli 119A, 00185 Roma, Italy.
[3]Deutsches Elektronen-Synchrotron DESY, Notkestraße 85, D-22607 Hamburg, Germany.
[4]Institute of Crystallography, CNR, via Salaria Km 29.300, 00015 Monterotondo Roma, Italy.
[5]Elettra Sincrotrone Trieste, Strada Statale 14-km 163,5, AREA Science Park, 34149 Basovizza, Trieste, Italy.

**Abstract**

Recently, x-ray illumination, using synchrotron radiation, has been used to manipulate defects, stimulate self-organization and to probe their structure. Here we explore a method of defect-engineering low-dimensional systems using focused laboratory-scale X-ray sources. We demonstrate an irreversible change in the conducting properties of the 2-dimensional electron gas at the interface between the complex oxide materials $LaAlO_3$ and $SrTiO_3$ by X-ray irradiation. The electrical resistance is monitored during exposure as the irradiated regions are driven into a high resistance state. Our results suggest attention shall be paid on electronic structure modification in X-ray spectroscopic studies and highlight large-area defect manipulation and direct device patterning as possible new fields of application for focused laboratory X-ray sources.

Author to whom correspondence should be addressed. Electronic mail:
a)n.poccia@utwente.nl
b)wang.xiao@utwente.nl

Intensive effort has focused on exploring the properties of low-dimensional materials systems in fundamental research and device engineering. A variety of materials have been studied: carbon based compounds, with graphene and carbon nanotubes being the most prominent representatives; semiconductors, like *e.g.* ultra-high-mobility GaAs heterostructures; and more exotic systems, such as the surface states of topological insulators, or the interfaces between complex oxides, which exhibit a particularly rich phase diagram[1]. As a direct result of low dimensionality, it is often possible to manipulate mechanical and electronic properties locally, and new functionality emerges. Recently, a *twin revolution* in X-ray optics and electron accelerator-based X-ray sources led to tremendous progress in the area of microfocusing, which allows to achieve high wavenumber resolution with micrometer-scale spatial resolution in X-ray microdiffraction experiments[2-4]. Moreover, with focal diameters of only a few hundred nanometers at synchrotron facilities[5], or tens of micrometers using improved laboratory X-ray sources, structural and electronic manipulation on a technologically relevant length scale is now possible[2-4,6-14]. In the process, crystallinity and phase-changes induced by X-rays can be monitored if high-resolution detectors are available, and real space imaging can be attained by scanning the focused X-ray beam over the sample[15-20]. With the use of laboratory sources, the X-ray fluence is easily adjusted by setting the anode current. This allows for accurate dosing which is an important feature for patterning applications.

The conducting interface between the complex oxide materials $SrTiO_3$ (STO) and $LaAlO_3$ (LAO)[21] is an interesting candidate system for electronic manipulation by hard X-rays. The material exhibits a thickness-dependent electronic reconstruction effect[22], but its conduction properties are also susceptible to oxygen-vacancy doping, lattice distortions, and substrate strain[1,23]. This results in a particularly rich phase diagram of electronic states at the interface: A high-mobility (quasi-) 2-dimensional electron gas (q2DEG) forms when the LAO thickness exceeds 3 unit cells (uc), resulting in an abrupt metal-to-insulator transition[22]. Superconductivity[24], magnetoresistance effects[25], and ferromagnetism[26] have been observed in different doping regimes. Achieving local control over oxygen vacancy sites in the vicinity of the interface[27–29], as well as over structural distortions[30] is highly desirable. It has been shown previously that such defects can be manipulated by synchrotron X-ray irradiation[2,3,9,10-13,]. Here, we utilize a laboratory focused X-ray source and study the regime in which smaller doses are applied in order to alter the conduction properties of the interface in the irradiated region. We map out a slow, irreversible increase in the resistance which saturates at >5 times its original value.

LAO/STO q2DEG samples were prepared by pulsed laser deposition of crystalline LAO on atomically flat, TiO$_2$-terminated STO (001) substrates. Because 10 uc of LAO provides the broadest coverage of intriguing properties, LAO is chosen to be 10 uc thick in this study. For electrical measurements, Hall bar devices were shaped by depositing a 10 nm-thick hard masking layer of amorphous LAO[31,32] prior to the growth of the crystalline material, which prevents the formation of a conducting interface in the masked regions. The substrates were treated for TiO$_2$-single termination using a standard process[33]. Using standard photolithography, we first defined a resist structure directly on top of STO surface. To prepare Hall bar devices, we employed the amorphous LAO (a-LAO) lift-off technique described by Schneider *et al*[34] and a 10 nm-thick a-LAO mask was deposited by pulsed laser deposition (PLD) from a single crystal target at room temperature in 2x10$^{-3}$ mbar oxygen. After lift-off patterning of the a-LAO, we grew crystalline LAO using PLD at the same oxygen pressure and a temperature of 850 °C. The deposition of this layer was monitored by *in-situ* reflection high-energy electron diffraction (RHEED), allowing growth control on a unit cell level (see Fig. 1a). After the growth the samples were annealed at 600 °C in 600 mbar O$_2$ for 1 hour. This ensured that the a-LAO/STO interface remained insulating. Finally, contact to the q2DEG was made by sputtering Ti/Au contacts at high-bias voltage. Figure 1b shows an optical image of the sample. The Hall bar devices are 85 µm wide and 1 mm long. The distance between voltage probes is 300 µm. The samples were mounted on a custom-made sample holder which connects to a Keithley 2400 SMU for resistance measurements. Figure 1c shows the setup configuration for the focused X-ray irradiation experiment.

The as-deposited devices showed typical sheet resistance values of 5 x 10$^4$ Ohm/square at room temperature. Electrical connections were made to a custom-made sample holder that was then installed into the X-ray system. The X-ray system is a PANalytical Empirean X-ray machine, equipped with a C-tech microfocus X-ray lens. The focal point of the X-ray beam was positioned at the center of the Hall bar using an optical microscope. The X-ray system has a 240 mm radius, 1.8 kW Cu line fine-focus X-ray tube. It is equipped with a strongly-focusing 50 µm polycapillary lens with a 5 µm pinhole as exit window and an intensity gain of ~10$^3$. The resulting focused X-ray beam has a Gaussian shape with a full width half maximum (FWHM) of about 60 µm and a photon flux of about 10$^6$ s$^{-1}$·µm$^{-2}$. The irradiated region covers 70 % of the width of the Hall bar as shown schematically in Fig. 1b that is approximately 1/6 of the device area. During irradiation, the device resistance was monitored using a Keithley 2400 Source-Meter Unit. The experiment was screened from ambient ultraviolet (UV) radiation. Before starting the first X-ray irradiation cycle, sufficient time elapsed for photoexcited carriers from natural light exposure to recombine.

In Figure 2a, we plot the room-temperature device resistance as a function of time measured for three consecutive cycles of X-ray illumination (cycle 1 from time $t_1$ to $t_3$, cycle 2 from $t_3$ to $t_5$, and cycle 3 from $t_5$ to $t_7$). The device resistance is the four-probe resistance (Ω/□) divided by the Hall bar width and the length between the voltage leads. All three cycles show qualitatively similar behavior: an initial sharp drop of the resistance when the X-ray beam is turned on, followed by a slow resistance increase. After the X-ray beam is switched off, the resistance rebounds and saturates at a higher value. As shown in the insert, similar effects were also seen in a Hall bar in a second sample. For the analysis, the impact of different cycles of X-ray exposition is shown for the first sample.

We attribute the initial resistance drop to photoexcited carriers[35-36] at the LAO/STO interface. Electrons are excited from the valence band (or subbands) to the conduction band of STO[37–39], perhaps with a minor contribution arising from electron excitation between STO in-gap states[40-42]. Photoconductivity contributes until the X-ray beam is turned off and all photoexcited carriers have recombined. Already during X-ray exposure, a slow irreversible resistance-change occurs to a value higher than that prior to illumination, indicating structural changes at the interface. It has been observed in synchrotron X-ray sources that the photons do more than generating carriers by exciting electron from the valence band to the conducting band[43,44]. In detail, the synchrotron X-rays produces band structure changes by directly and/or indirectly alternating the Ti-O buckling or other ordering of the domain structure in the interface region of STO[30,43,44]. Therefore we propose that the irreversible resistance-change is possibly due to the same structural effect. For a quantitative assessment, we fit the measured resistance in the time intervals from $t_1$ ($t_3$, $t_5$) to $t_2$ ($t_4$, $t_6$), respectively, to a double exponential decay function,

$$R(t) = c_1 \times e^{-(\frac{t}{\tau_1})} + c_2 \times e^{-(\frac{t}{\tau_2})} + R_0$$

where the prefactors $c_1$ and $c_2$ are constants, and $R_0$ is the initial device resistance of the Hall bar at each cycle. When the X-ray beam is switched off (at $t_2$, $t_4$, and $t_6$), the photoexcited electron-hole pairs recombine. The resistance can be fit by the well-known Kohlrausch expression[45–47]:

$$R(t) = c_3 \times e^{-(\frac{t}{\tau_3})^\beta} + R_0$$

where $c_3$ depends on the recombination process and $R_0$ is the device resistance at each cycle when X-ray is switched off. Figure 2b shows that, between cycles 1 and 3, $\tau_3$ had decreased from about 1000 seconds to 800 seconds whereas $\beta \approx 0.5$ throughout. The time constants $\tau_3$ and $\beta$ are comparable to reported values in UV photoconductivity studies[36,38]. Figure 3 shows the fitting accuracy and fitted curves for the excitation and relaxation processes in the different cycles. All curved are fitted by the two formulas, indicating the feasibility of the two models.

The fitted parameters are summarized in the figure 4. Figure 4a shows trends for three consecutive irradiation cycles: $\tau_1$ increased from less than 4 seconds to above 45 seconds, indicating that it becomes more and more difficult to photoexcite carriers. $\tau_2$, on the other hand, dropped from an initial value of 550 s by roughly a factor of 2. The focused X-rays induces band structure changes by creating structural changes. For a fixed flux per unit area, only a finite density of structural changes can be generated. This is reflected in the moderate positive resistance slope and in its tendency to flatten in time. During the 4 minutes, before the X-ray beam is switched off at $t_6$, the resistance is almost constant. This also explains the increasing difficulty on generating structure changes at t a fixed flux and on photoexciting carriers with increasing of focused X-ray irradiation time. Therefore the evolution of the exponents indicates an effect of the micro X-rays on the structural changes and on the electronic states at the interface. Furthermore, the value $\beta=\delta/(\delta+2)=0.5$ indicates glassy behavior in a ($\delta=2$)-dimensional system[48,49] compatible with the 2D nature of conductivity[21] and electronic[26] phase separation[26] at the interface. It is worth to observe that this phase separation could be of the same nature of the one observed in cuprates[50,51].

For the 10 uc LAO/STO interfaces, the induced high-resistance state is stable (tested >16 hours), indicating a permanent change of the interface electronic states. A lower-bound estimate of the change in sheet resistance for the exposed area yields a more than fivefold increase compared to the resistance prior to X-ray exposure. When LAO thickness is greater than 4 uc, all the LAO/STO interfaces share the same nature of electronic structure and thus the manipulation of electronic state is expected to validate in all the conducting LAO/STO interfaces. When LAO thickness is smaller than 4 uc, the LAO/STO interface abruptly switches to an insulating state[22]. In this insulating case, the focused micro X-ray should only preserve the insulating electronic state.

The observed interaction between focused micro X-ray and low dimensional oxides is in particular significant for fundamental X-ray spectroscopic studies and potential technological applications. The widely used X-ray spectroscopic study, such as X-ray absorption spectroscopy and angle resolved photoemission spectroscopy *etc.*, is one of the most important techniques for probing electronic structures in modern condense matter physics. Our observation points out that extra attention shall be put on the X-ray induced permanent electronic structure changes when conducting X-ray spectroscopic experiments and interpreting the research data. Secondly, this result also indicates the feasibility of using laboratory X-ray sources for direct device patterning in this and similar systems with further developments in the equipment aimed to higher X-rays flux. Comparing with other aggressive techniques, such as ion implantation, masked chemical, plasma treating, sputtering or simple electron beam bombardment, our technique is more modest and offers possibility of tuning the electronic resistivity to a desired resistivity state. This is also attractive for other electronic structure related applications, such as electrochemistry reactions and sensing.

In conclusion, using focused X-rays, we have manipulated the electrical properties of LAO/STO interfaces. Upon X-ray irradiation, a photoconductivity process and an X-ray induced permanent resistance change occur. The resistance increase is analyzed in detail and correlated with the evolution of the time constants under several cycles of irradiation. From an initial value, we can tune up the resistance to desired values in a permanent fashion. This indicates existence of electronic structure modification in X-ray spectroscopic studies and provides additional insights into data interpretation of the spectroscopic results. Furthermore, our report, together with the latest technological developments in laboratory X-ray sources and optics, opens a possibility to control and modify sensitive materials in-house with less expensive and space consuming laboratory sources as compared to the synchrotron sources that were mostly used for this purpose until now.

**Acknowledgements:** We thank the PANalytical company for providing us with access to the focused X-ray source. We are grateful to Alexander Kharchenko, Detlef Beckers, Eugene Reuvekamp, Joachim Woitok and Martijn Fransen at PANalytical for their support. We thank the ELETTRA and the European Synchrotron Radiation Facility staff for support. We thank Frank Roesthuis and Dick Veldhuis for help and support during the experiments and Alexander Brinkman and Gertjan Koster for useful discussions. The work was supported by the Dutch FOM and NWO foundations. N.P. acknowledges for financial support the Marie Curie IEF project for career development. X.R.W. acknowledges for financial support the Dutch Rubicon grant.

**Competing Interests:** The authors declare that they have no competing financial interests.

**Figures and figures captions:**

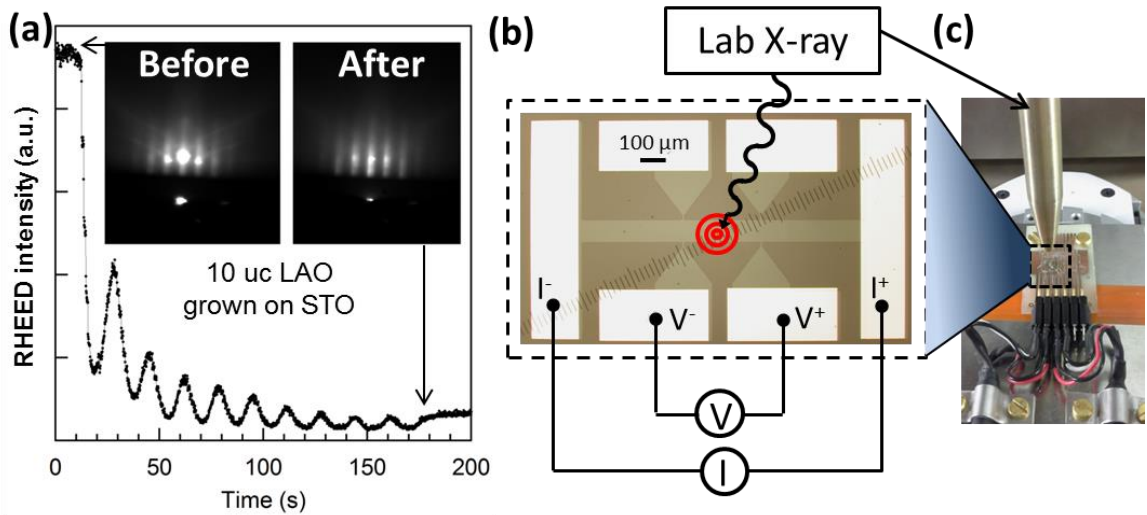

**Figure 1. Device fabrication and measurement setup. (a)** Real-time intensity of the specular reflected RHEED spots during deposition of a ten unit cell-thick layer of LaAlO$_3$, at 850 ℃. In the left and right insets, the RHEED diffraction patterns of SrTiO$_3$ at 850 ℃ before and after the 10 uc LaAlO$_3$ deposition are shown respectively. **(b)** Optical image of the Hall bar device, structured as described in the main text. **(c)** A close-up image of the experimental setup. The X-ray source is on top (not depicted), and the exit window is visible. The sample position can be adjusted in x-, y-, and z-directions. The resistance is measured at room temperature during the X-ray illumination at the indicated spot.

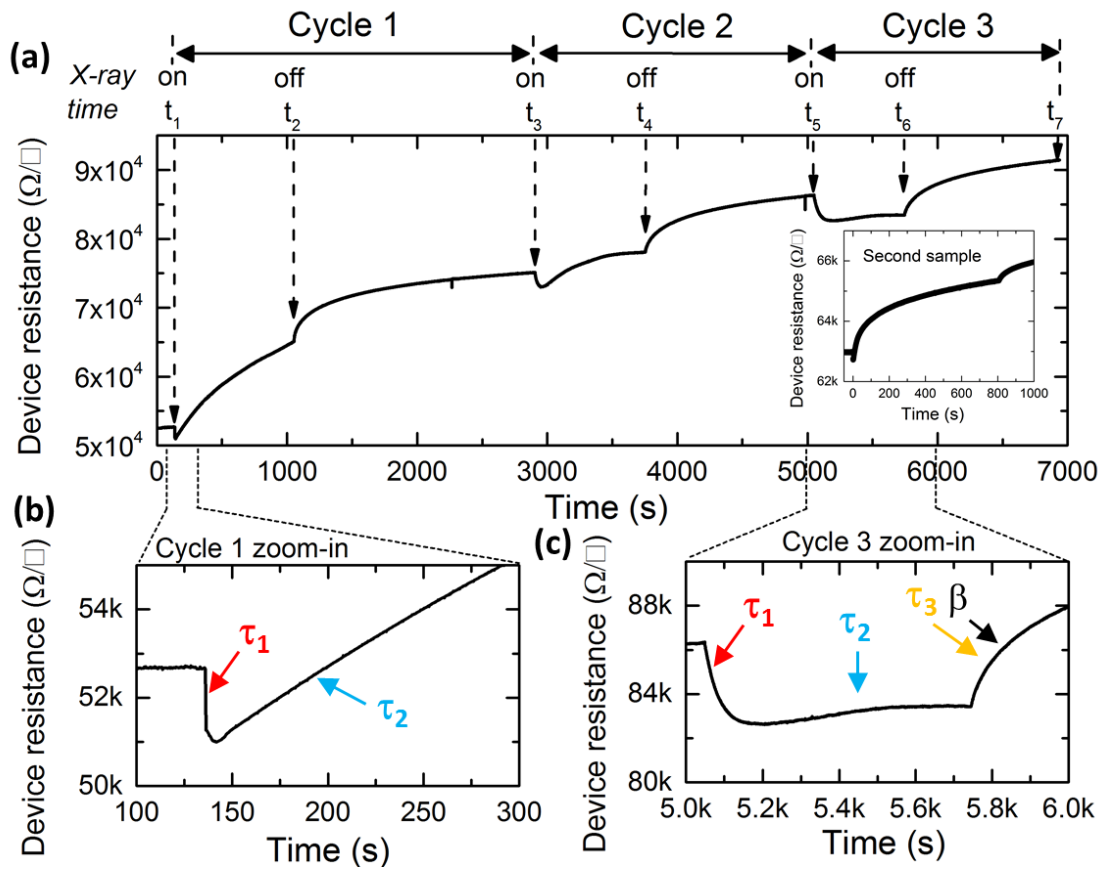

**Figure 2. Device resistance as a function of irradiation time with the PANalytical focused X-ray source.** **(a)** The device resistance for 3 cycles (on/off) of X-ray irradiation. Similar effects observed in the second sample are shown in the insert. **(b)** Detail of resistance vs. time curve of cycle 1. **(c)** Detail of resistance vs. time curve of cycle 3. Three characteristic processes with different time constants are observed: excitation of photocarriers ($\tau_1$), resistance increase due to irreversible structural changes ($\tau_2$) and recombination of photocarriers after the X-ray source is switched off ($\tau_3$).

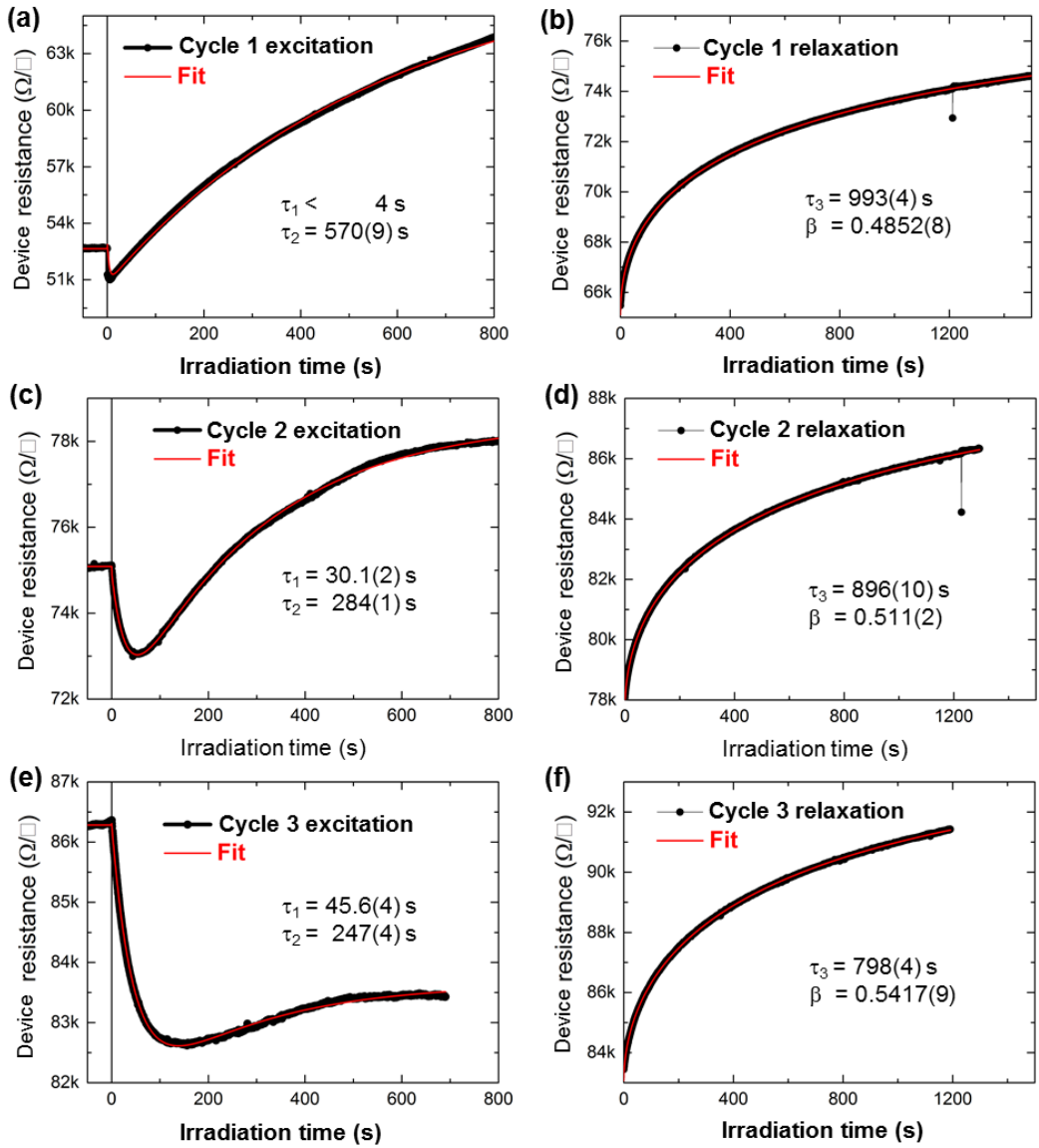

**Figure 3. Time constants used for fitting different process of three consecutive cycles.** Double exponential fit for the cycle 1 in (a), 2 in (c), and 3 in (e). Kohlrausch fit for the relaxation process of cycle 1 in (b), cycle 2 in (d) and cycle 3 in (e).

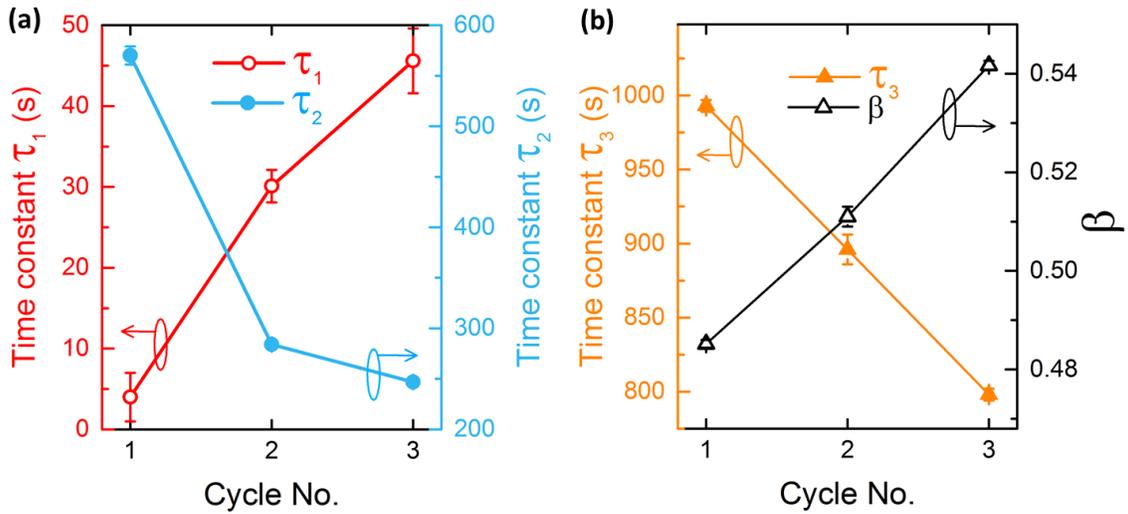

**Figure 4. Time constants for the resistance fits of three consecutive irradiation cycles.** (**a**) The time constants for photoconductivity and structural changes increase and decrease, respectively, with each additional cycle. (**b**) The evolution of parameters $\tau_3$ and $\beta$ in the stretched exponential function used for fitting the relaxation process.